\documentclass[aps,pra,twocolumn,showpacs,preprintnumbers]{revtex4}
\usepackage{graphicx}
\usepackage{dcolumn}
\usepackage{bm}

\begin{document}

\title{Observation of multiple ionization pathways for OCS in an intense
laser field resolved by three dimensional covariance mapping and visualized
by hierarchical ionization topology.}
\author{W. A. Bryan}
\email{w.bryan@ucl.ac.uk}
\author{W. R. Newell}
\affiliation{Department of Physics and Astronomy, University College London, Gower
Street, London WC1E 6BT, UK}
\author{J. H. Sanderson}
\affiliation{Department of Physics, University of Waterloo,
Waterloo, Ontario, Canada N2L 3G1}
\author{A. J. Langley}
\affiliation{Central Laser Facility, Rutherford Appleton Laboratory, Chilton, Didcot,
Oxon OX11 0QX, UK}
\date{\today }

\begin{abstract}
The two- and three-body Coulomb explosion of carbonyl sulfide (OCS)
by 790 nm, 50 fs laser pulses focussed to $\approx $ 10$^{16}$
Wcm$^{-2}$ has been investigated by three-dimensional covariance
mapping technique. For the first time in a triatomic molecule, a
single charge state, in this case the trication, has been observed
to dissociate into two distinct energy channels. With the aid of a
three dimensional visualization technique to reveal the ionization
hierarchy, evidence is presented for the existence of two sets of
ionization pathways resulting from these two initial states. While
one group of ions can be modeled using a Classical enhanced
ionization model, the second group, consisting of mainly asymmetric
channels, can not. The results provide clear evidence that an
enhanced ionization approach must also be accompanied by an
appreciation of the effects of excited ionic states and
multi-electronic processes.
\end{abstract}

\maketitle



\section{Introduction}

Advances in tabletop femtosecond laser generation and amplification
techniques has lead to the routine availability of light pulses
which may be focussed to intensities comparable to the binding
between the proton and electron in atomic hydrogen \cite{ref1}. When
a triatomic molecule is exposed to such a light pulse, the dynamics of
electron removal and subsequent Coulomb explosion is rich in its
complexity and not yet fully understood. Typically, the variation of
the focussed laser intensity
throughout the confocal volume \cite{ref2,ref3} exposes the molecule to $%
\simeq $ 10$^{16}$ Wcm$^{-2}$ at focus, decreasing to such an extent that in
the low intensity lobes before and after the focus that the effects are
purely perturbative. Depending on the location and orientation of the
molecule, the laser pulse can initiate a number of complex processes,
including laser-induced reorientation \cite{ref4}, low energy dissociation
with the removal of 0, 1 or 2 electrons \cite{ref4,ref5} and two- or
three-body Coulomb explosion (2BCE and 3BCE respectively) \cite%
{ref4,ref6,ref7,ref8,ref9,ref10,ref11,ref12,ref13,ref14,ref15,ref16} with
the removal of at least 3 electrons. Additional complexity is introduced
through the observation that the laser pulse may modify the geometry of
triatomics undergoing 3BCE (H$_{2}$O \cite{ref6}, OCS \cite{ref7}, CO$_{2}$ %
\cite{ref4,ref8,ref9,ref14,ref15,ref16}, SO$_{2}$ \cite{ref10}, NO$_{2}$ %
\cite{ref11}). \ The recent development of complete momentum measuring
coincidence techniques, \cite{refC} and the advent of sub 10 femtosecond
laser pulses has allowed the observation of the near equilibrium geometry of
triatomic molecules through Coulomb imaging \cite{refB}. \ In addiction both
concerted and sequential dissociation channels have been mapped in time for
SO$_{2}^{2+}$\cite{refA}$.$Concerted and sequential channels in the triply
charged ions of N$_{2}$O and CS$_{2}$ \cite{refD, refE, refF} have also been
identified. 3BCE is further complicated in many-electron triatomics by the
large number of possible ionization channels, here labelled according to the
convention OCS$^{(m+n+p)+}$ $\rightarrow $ O$^{m+}$ + C$^{n+}$ + S$^{p+}$
and referred to as the ($\mathit{m}$, $\mathit{n}$, $\mathit{p}$) channel.

Theoretical treatments of two- and three-body Coulomb explosion have
been proposed and may be separated into quantum mechanical
\cite{ref17,ref18} and classical \cite{ref19} interpretations,
broadly termed enhanced ionization (EI). However recent experimental
and theoretical works present possible shortcomings. A number of
theoretical considerations indicate that molecular ionization cannot
be treated as the action of a single electron. Specifically,
\textit{ab initio} calculations of the distortion of carbon dioxide
during multiple ionization by a static electric field \cite{ref20}
indicate the presence of a charge-exchange mechanism leading to
experimentally observed geometries \cite{ref4,ref9} and ionization
proceeding through the most negative atomic site. The ongoing
theoretical investigation into the
applicability of Thomas-Fermi theory to ultrafast ionization in molecules %
\cite{ref21} has lead to a recent publication concerning the ionization of CO%
$_{2}$ and N$_{2}$O\cite{ref22}. This hydrodynamic treatment of a
group of active electrons has resulted in an impressive agreement
between experiment and calculation, however the extent to which the
predicted dynamics are physical is open to doubt. \ This is
particularly clear in the case of CI$_{2}$ \cite{ref21} where the
model generates a CI-CI bond much stiffer than is natural which
subsequently plays an important part in keeping the molecule at low
bond length as ionization proceeds. \

An important consideration in molecular ionization is the sequence
of events within the laser pulse temporal envelope. A strong
coupling exists between electron removal and the expansion of the
molecular bonds. Thus, as the molecule is ionized - either
sequentially or non-sequentially, the rate of expansion of the
molecular bonds can vary, strongly influencing further ionization.
As proposed in enhanced ionization \cite{ref19}, a classical
treatment of molecular ionization by Posthumus and coworkers,
multiple ionization is governed by dissociation, which gives rise to
sequential ionization every time a classical ionization threshold is
crossed. The 'critical separation' in the bond length, where the
final ionization takes place, is simply the consequence of the
strong dip in the threshold intensity dependence on bond length.
Typically, these critical separations are two or three times the
ground-state equilibrium bond length depending on the molecular
type. The threshold dip is the consequence of the combined effect of
electron localization and Stark shifting. We have previously shown
that the adaptation of this method to triatomics such as OCS can
give quantitative agreement with experiment. A criticism of both the
classical enhanced ionization treatment and hydrodynamic models is
that neither makes any account the nonadiabatic response of a group
of electrons to a laser pulse of a specific frequency and intensity,
which has been shown to be influential \cite{ref23}. Exact numerical
solutions of the hydrogen molecular ion by Bandrauk and co-workers
\cite{ref17} have independently predicted a maximum in the
ionization rate through a different mechanism. The electric field
generated by the laser pulse produces a Stark-shift of the
lowest-unoccupied molecular orbital (LUMO). For particular
internuclear separations, this Stark-shifting lifts the LUMO above
the Coulomb barrier. However, for ionization to proceed, the LUMO
must be populated: as the molecule expands during the laser pulse,
the exchange of population between the highest occupied molecular
orbital (HOMO) and the LUMO becomes nonadiabatic, producing a net
population in the LUMO. Furthermore, Kawata, Kono and Bandrauk have
predicted similar maxima in the ionization rate in the case of
linear H$_{3}^{+}$ in an intense laser field \cite{ref24}. As in the
case of H$_{2}^{+}$ \cite{ref17}, the response of the molecule to
the laser field can be classified as either adiabatic or
nonadiabatic. In the adiabatic system, electron transfer along the
molecule is possible, whereas in the nonadiabatic case electron
migration is suppressed due to laser induced localization. The
ionization dynamics of this two-electron system depend heavily on
both the electron distribution along the molecule and the
nonadiabatic transitions between the lowest three electronic states.

\begin{figure}[tbp]
\includegraphics[width=235pt]{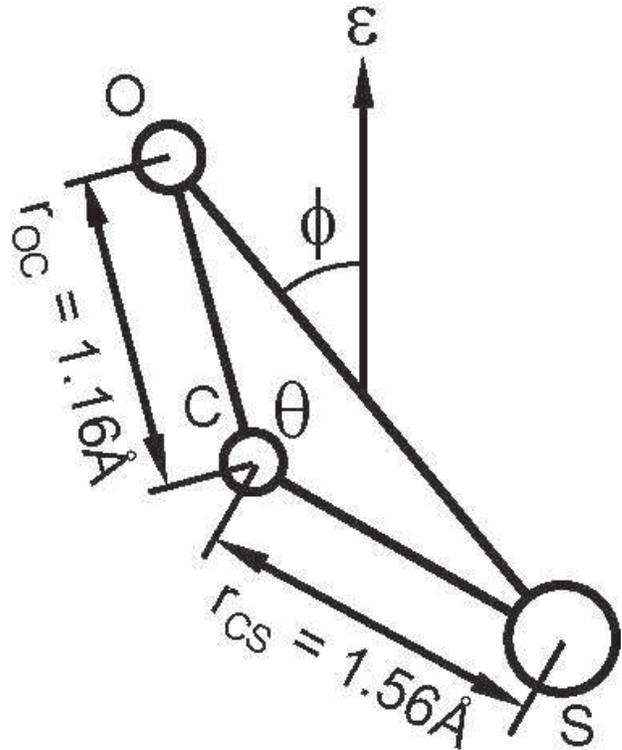}
\caption{Schematic of the carbonyl sulfide molecule currently under
investigation.}
\label{fig1}
\end{figure}

These processes although not explicitly predicted for multi-electron systems
have been the subject of some detailed experimental work \cite{refG, refH}%
. By making a careful examination of the bond lengths at which channels of
the exploding N$_{2}$ and I$_{2}$ molecules appear to occur, it was deduced
that certain channels could not be precursors of others and that because of
this it was not possible that the symmetric enhanced ionization channel
(1,2)-(2,2) could exist. It was therefore necessary to invoke
multi-electronic processes causing asymmetric ionization pathways ie
(1,2)-(1,3). The authors postulate that this process still involves the
strong coupling of $g-u$ symmetry states (analogous to HOMO, LUMO in the
enhanced ionization symmetric ionization process) but that the strong
coupling of the outer electron to the field gives rise to the ionization of
an inner electron (internal re-scattering). Such multi-electronic processes
would leave the molecular ion in excited states and direct evidence of this
has been found in fluorescence studies of N$_{2}$ \cite{refJ, refN} and
various other molecules \cite{refI}. One innovative features of \cite{refH}
was to introduce a visualize method able to show the progression of
ionization through many available pathways.

In the case of a near-linear \cite{ref7} triatomic such as OCS, the
molecular geometry is defined by the bond lengths
(\textit{r}$_{\mathrm{OC}}$ and \textit{r}$_{\mathrm{CS}}$) and the
bend angle ($\theta $) between the bonds, as defined in figure 1,
these are the controlling factors in EI. The current work continues
our investigation of the behaviour of triatomics in a laser field;
in the current work, we return to the carbonyl sulfide (OCS)
molecule to resolve the ionization sequence. Our previous work
\cite{ref7} used momentum mapping and classical enhanced ionization
to resolve the bond lengths at ionization for the 334 channel. We
were able to show that the difference between the bond length
calculated using a simplistic Coulomb assumption (the molecule is
stationary before explosion) is within ten percent of the correct
bond length calculated using an iterative technique.
Although ionization channels have been identified in CO$_{2}$ \cite%
{ref4,ref12,ref13,ref15}, and the possibility of non-sequential ionization
taking place in order to reach the (111) channel has been discussed \cite%
{refM} the transitions between the higher channels have not been probed. By
choosing the highly asymmetric OCS molecule instead of CO$_{2}$ we might
expect to be able to find the limits of applicability of the classical
ionization approach. Where possible our aim is to clarify the role of
transitions between channels through the visualization of a hierarchical
ionization topology.

The covariance mapping (CM) technique was used to investigate laser-molecule
interactions as early as 1989 \cite{ref25}, but it was 1994 before
exhaustive investigations into ionization in CO$_{2}$ were published.
Frasinski \textit{et al} \cite{ref12} employed the technique to identify
ionization channels; the observed channels were confirmed by Cornaggia
\textit{et al} \cite{ref13} using a variant of the technique. Cornaggia
performed further experiments on polyatomic molecules using CM, reporting a
straightening of carbon dioxide in 130 fs laser pulses \cite{ref14}. Without
implying any compromise in the conclusions published by Cornaggia and
co-workers, it has since become apparent that the CM technique is not
ideally suited to investigating the geometry of exploding triatomics, which
requires unit collection efficiency for all ions, irrespective of momentum.
This situation was addressed initially by Hishikawa and co-workers who
proposed the mass-resolved momentum imaging (MRMI) technique \cite%
{ref9,ref10,ref11}. An analogue to this technique, ion-momentum imaging
(IMI), was developed by the authors and co-workers, and applied to the 3BCE
of H$_{2}$O \cite{ref6}, CO$_{2}$ \cite{ref4} and OCS \cite{ref7} where
laser-induced geometry modifications were reported and compared to Monte
Carlo simulations. Importantly, the IMI technique also includes an
instrumental correction process \cite{ref4,ref6}, that removes the
instrumental bias by quantifying the variation of detector efficiency with
ion species and momentum. W T Hill and co-workers examined molecular
distortion by isolating 3BCE products from a specific channel and bend
angle, then making comparisons between the EI and the Thomas-Fermi models %
\cite{ref16}. This novel approach indicates the Thomas-Fermi and EI schemes
may be linked via an effective charge defect.

In this paper, dissociation and 2BCE are investigated using two-dimensional
covariance mapping (2DCM). Measurements of the kinetic energy of the product
ions is reported. The 2DCM technique has previously been used to investigate
3BCE of CO$_{2}$ \cite{ref4} and OCS \cite{ref7}, however, given that OCS is
comprised of three different atomic species, and that the charge-to-mass
ratio of oxygen and sulphur atoms is degenerate under certain ionization
states, three-dimensional covariance mapping (3DCM) of the Coulomb explosion
of OCS is necessarily employed in the presented work. This allows
unambiguous identification of the ionisation channels and measurement of
their relative strengths. The kinetic energy release associated with each of
these channels is also reported, allowing the OC and CS bonds to be
measured. This paper is organised as follows: the experimental configuration
will be discussed in the following section, with details of the data
collection and processing. Results of 2DCM experiments will then be
presented, and details of the observed OCS 2BCE channels introduced. The
3DCM of OCS will then be presented, with particular discussion concerning
the application of modern visualisation techniques, allowing new insights
into the behaviour of a triatomic during Coulomb explosion. Measurements of
the r$_{OC}$ and r$_{CS}$ bond lengths will then be presented for each
channel, along with an estimate of the relative channel strengths. At this
point, the scheme of hierarchical ionization topology will be introduced,
and the 3DCM results reinterpreted using a 3D diagrammatic technique adapted
from \cite{refH}.

\section{Experimental configuration and Data processing}

As in previous experimental studies, the RAL (UK) ASTRA laser facility
generated the femtosecond laser pulses, as detailed in previous publications %
\cite{ref4,ref6,ref7,ref8}. Ti: Sapphire seed pulses at 790 nm were
amplified by CPA \cite{ref26} to a pulse energy of 1 mJ in 50 fs at a
repetition rate of 10 Hz. An f/5 focus was then generated in the source
region of a Wiley-McLaren \cite{ref27} time-of-flight mass spectrometer
(TOFMS), interacting the laser pulses with the gas target. The background
pressure in the TOFMS system is $\simeq$ 8 $\times$ 10$^{-10}$ mbar. Product
ions generated by the interaction of the focused laser pulse with the target
gas were extracted from the source region, and detected using a pair of
micro-channel plates after 11 cm of field-free drift. Throughout the present
work, the laser polarization direction is parallel to the axis of the TOFMS.
Unlike previous IMI experiments, the detector is now exposed to the full
confocal volume and operated in high efficiency mode, with an extraction
field of 600 Vcm$^{-1}$. The output of the TOFMS is monitored on a Tektronix
TDS-744A digital storage oscilloscope (DSO). Time-of-flight spectra were
recorded on the DSO, and stored on a laboratory PC for off-line analysis.
The two- and three-dimensional covariance coefficients (C$_{2}$($\mathit{x}$%
, $\mathit{y}$) and C$_{3}$($\mathit{x}$, $\mathit{y}$, $\mathit{z}$)
respectively) were then evaluated for all $\mathit{n}$ points for each of
the $\mathit{N}$ TOF spectra. For definitions of C$_{2}$($\mathit{x}$, $%
\mathit{y}$) and C$_{3}$($\mathit{x}$, $\mathit{y}$, $\mathit{z}$), the
reader is referred to reference \cite{ref28}, which provide full discussions
of the application of covariance mapping to molecular ionization. Depending
on the process under investigation, either C$_{2}$($\mathit{x}$, $\mathit{y}$%
) (2BCE) or C$_{3}$($\mathit{x}$, $\mathit{y}$, $\mathit{z}$) (3BCE) was
evaluated for all $\mathit{n}$ points on the TOF spectrum, resulting in an $%
\mathit{n}^{2}$ matrix in the case of C$_{2}$($\mathit{x}$, $\mathit{y}$),
referred to as the two-dimensional covariance map (2DCM) and an $\mathit{n}%
^{3}$ matrix in the case of C$_{3}$($\mathit{x}$, $\mathit{y}$, $\mathit{z}$%
), similarly referred to as the three-dimensional covariance map (3DCM).
Throughout this paper, the axes $\mathit{x}$, $\mathit{y}$ are defined as
the time-of-flight axes in the 2DCM, and $\mathit{x}$, $\mathit{y}$ and $%
\mathit{z}$ are defined as the time-of-flight axes in the 3DCM.

To test the performance of the experimental apparatus with respect to false
correlations \cite{ref28}, a series of two-dimensional maps were recorded
for the order of 3000 laser shots as a function of target gas pressure. At a
target gas pressure of 3 $\times$ 10$^{-9}$ mbar, the contribution of false
correlations to the covariance map became negligible. Of the order of 10$%
^{5} $ individual TOF spectra were recorded directly to hard disk.

\section{Results - 2BCE of OCS}

\begin{figure}[tbp]
\includegraphics[width=235pt]{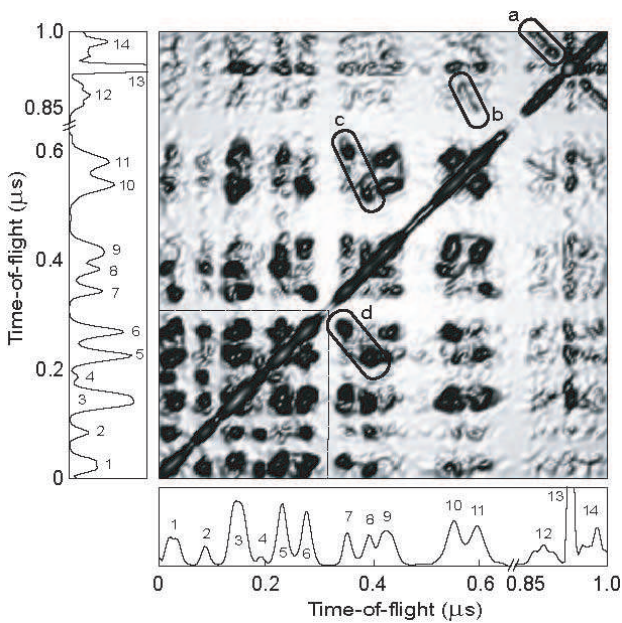}
\caption{Two-dimensional covariance map of carbonyl sulfide recorded with
790 nm 50 fs Ti: Sapphire laser pulses focused to an intensity of $\simeq $
10$^{16}$ Wcm$^{-2}$. The regions of positive covariance are identified by
the labelled average time-of-flight spectra, the peaks are identified in
table 1. The features identified $\mathit{a}$ - $\mathit{d}$ are discussed
in the text.}
\label{fig2}
\end{figure}

Figure 2 is a surface representation of the 2DCM of OCS after
10$^{4}$ laser shots, under the optical conditions discussed in the
previous section. Correlations on the map are identified by labeled
average time-of-flight spectra to the left and below the map, where
the peaks in the ionization signal are identified in table I. As is
apparent from table I, peaks 3, 5, 6, 10 and 11 are the result of
the detection of two or more ions, either through charge degeneracy
(peaks 5, 6, 10 and 11) or temporal overlap. As a consequence, the
average time-of-flight spectrum gives a limited amount of
information about the explosion of the molecule. Note that both TOF
axes are truncated between peaks 11 and 12, as no ions are detected
between t = 650 and 850 ns. The major features of figure 2 are as
follows. The strong
diagonal signal is the autocorrelation line, a consequence of an event at $%
\mathit{x}$ = $\mathit{y}$ ~always being self-correlated and therefore gives
no information about the processes under investigation. True correlations
lie above and below this line, false correlations associated with peak 13
(OCS$^{2+}$) are unavoidable. The double ionization of the molecular ion
requires a lower intensity as compared to the multiple ionization leading to
Coulomb explosion, hence the volume of the focus generating OCS$^{2+}$ is
approximately an order of magnitude greater than that generating the Coulomb
explosion signal \cite{ref2}.

\begin{table}[tbp]
\caption{Peak assignments for the average time- of-flight spectra presented
in the present work}
\label{tab:table1}%
\begin{ruledtabular}
\begin{tabular}{lc}
Peak &Ion Species\\
\hline
1 &C$^{3+}$\\
2 &O$^{3+}_{F}$\\
3 &C$^{2+}$ / O$^{3+}_{B}$ / S$^{5+}_{F}$\\
4 &S$^{5+}_{B}$\\
5 &O$^{2+}_{F}$ / S$^{4+}_{F}$\\
6 &O$^{2+}_{B}$ / S$^{4+}_{B}$\\
7 &S$^{3+}_{F}$\\
8 &S$^{3+}_{B}$\\
9 &C$^{+}$\\
10 &O$^{+}_{F}$ / S$^{2+}_{F}$\\
11 &O$^{+}_{B}$ / S$^{2+}_{B}$\\
12 &CO$^{+}_{F/M/B}$\\
13 &OCS$^{2+}$\\
14 &S$^{+}_{F/B}$\\
\end{tabular}
\end{ruledtabular}
\end{table}

Important features of figure 2 are highlighted a - d, and are identified
thus: (a) CO$^{+}$ + S$^{+}$, (b) CO$^{+}$ + S$^{2+}$, (c) S$^{3+}$ + O$^{+}$
and (d) S$^{3+}$ + O$^{2+}$. Correlations (a) and (b) identify the 2BCE
channels (1, 1) and (1, 2) respectively. The kinetic energy release (KER)
associated with these 2BCE processes, channels a and b, may be calculated
from the temporal width in $\mathit{x}$ and $\mathit{y}$ of the
correlations, and are presented in table II. These channels, (1,1) and (1,2)
are expected in the 2DCM, as the OC bond is considerably stronger than the
CS bond, as observed in recent dissociative photoionization experiments
performed by Eland and co-workers \cite{ref29} and Brion and co-workers \cite%
{ref30}, implying that early ionization and dissociation favour
energetically efficient routes. Wang and Vidal \cite{ref31} have published
cross sections for the electron impact dissociative ionization of OCS, and
reported the CO$^{+}$ + S$^{+}$ channel has a cross section at least five
times that of any channel involving CS$^{+}$. In the present work, only a
trace of CS$^{+}$ is detected at a longer flight time than that presented in
figure 2, however it is uncorrelated on an equivalent 2DCM, implying O$^{+}$
+ CS$^{+}$ is negligible, and the CS$^{+}$ signal probably originates from
the very weak O + CS$^{+}$ dissociative process.

Because of the charge degeneracy and temporal overlap present in much of the
average time-of-flight spectrum, the dense area of the 2DCM containing the
atomic product ions where \textit{t} $\leq $ 0.65 $\mu s$ is difficult to
interpret. However, relying on the fact that peaks 7 and 8 are purely due to
S$^{3+}$, correlations (c) and (d) can be employed to make limited
inferences about the molecular behaviour. Correlations (c), peaks 7 (S$%
_{F}^{3+}$) and 11 (O$_{B}^{+}$ / S$_{B}^{2+}$) peaks 8 (S$_{B}^{3+}$) and
10 (O$_{F}^{+}$ / S$_{F}^{2+}$) are forwards - backwards pairs, formed by
the ions being produced parallel to the laser polarization direction.
Similarly in correlation (d) peaks 7 (S$_{F}^{3+}$) and 6 (O$_{B}^{2+}$/S$%
_{B}^{4+}$) and peak 8 (S$_{B}^{3+}$) and 5 (O$_{F}^{2+}$/S$_{F}^{4+}$) are
forward backward pairs. As observed in \cite{ref7} the molecule favours
exploding with the O - S axis aligned to the laser field, i.e. $\varphi $ $=$%
0 in figure 1. All triatomic molecules exposed to similar laser conditions
have been found to exhibit this forward-backward behaviour.
However other correlations are overlapping such as those associated with
peak 3 which represents C$^{2+}$ as well as O$_{b}^{3+}$ and S$_{f}^{5+}$.

\section{Results - 3BCE of OCS}

To establish the 3BCE channels, we turn to the 3DCM results. Due to the
extra dimension introduced into the calculations, the correlation islands on
the 2DCM, figure 2, become correlation volumes in the 3DCM. True 3BCE
correlations appear as small volumes of positive covariance within the map
volume. Identifying the locations of these correlations is complicated by
the presence of three autocorrelation (AC) planes, which traverse the map at $%
\mathit{x}$ = $\mathit{y}$, $\mathit{x}$ = $\mathit{z}$ and $\mathit{y}$ = $%
\mathit{z}$. As with the AC line in a 2DCM, volumes of the 3DCM located on
the AC planes have a stronger covariance than 3BCE correlations but are
essentially meaningless. The geometry of the AC planes is illustrated in
figure 3, and will be present in any 3DCM.

\begin{figure}[tbp]
\includegraphics[width=235pt]{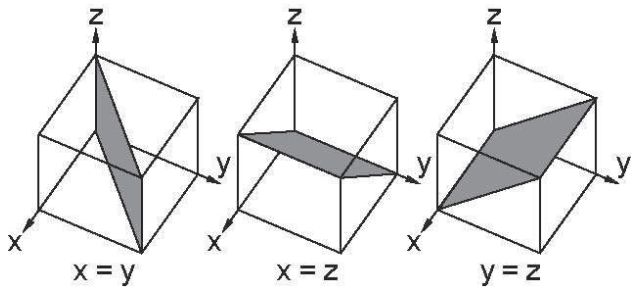}
\caption{Geometry of the autocorrelation planes present in a
three-dimensional covariance map}
\label{fig3}
\end{figure}

Research Systems IDL (Interactive Data Language) 5.5 was used to examine the
OCS 3DCM, allowing us to present, for the first time, visualizations through
the body of the 3DCM. Previous studies \cite{ref12,ref13,ref15,ref28}
present two-dimensional slices through the body of the 3DCM, where a slice
or plane parallel to the $\mathit{x}$-axis shows all correlations in $%
\mathit{y}$ and $\mathit{z}$. While being a reasonable manner in which to
view a complex dataset, when viewed in three-dimensions, the location of
local maxima may be identified more easily and accurately. A voxel (volume
element) projection of the 3DCM between 0 $\leq $ ($\mathit{x}$, $\mathit{y}$%
, $\mathit{z}$) $\leq $ 300 ns is presented in figure 4, corresponding to
the region in figure 2 between the origin and the dashed lines. The full
3DCM was calculated up to \textit{t} = 1 ms, but for the purposes of
presentation, only a limited subset of this volume presented in figure 4.

\begin{figure}[tbp]
\includegraphics[width=235pt]{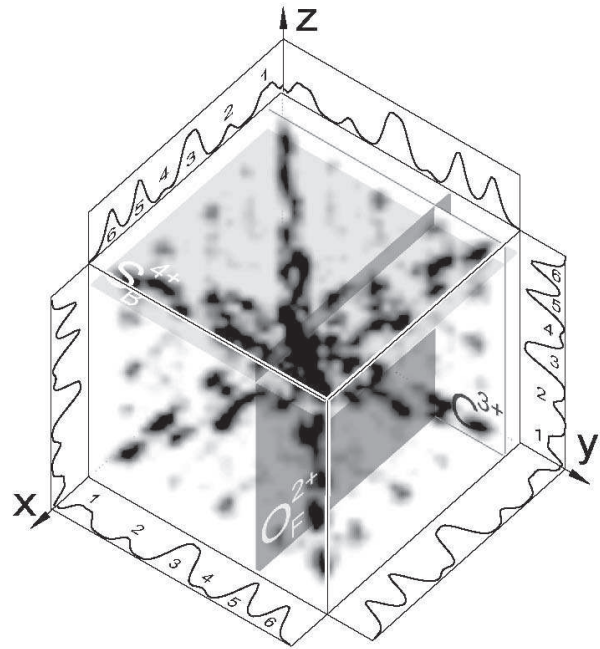}
\caption{Projection visualization of the OCS three-dimensional covariance
map. The range in ($\mathit{x}$, $\mathit{y}$, $\mathit{z}$) corresponds to
the region within the dashed box in figure 2. The peaks are identified by
comparison with table I. Three semitransparent planes illustrate the
location of the (2, 3, 4) channel.}

\label{fig4}
\end{figure}

The visualization presented in figure 4 is generated in two steps. Firstly a
transparency threshold, $\mathit{T}$ ~is defined such that if C$_{3}$($%
\mathit{x}$, $\mathit{y}$, $\mathit{z}$) $\leq$ $\mathit{T}$, the point ($%
\mathit{x}$, $\mathit{y}$, $\mathit{z}$) is treated as transparent. If C$%
_{3} $($\mathit{x}$, $\mathit{y}$, $\mathit{z}$) $>$ $\mathit{T}$, the point
($\mathit{x}$, $\mathit{y}$, $\mathit{z}$) is assigned a variable opacity,
where the opacity depends logarithmically on C$_{3}$($\mathit{x}$, $\mathit{y%
}$, $\mathit{z}$). Secondly a three-dimensional transform is applied
allowing arbitrary rotation, translation and oblique display of the 3DCM
onto an image plane, in this case defined as the plane of the figure.
Visualization in this manner is similar to taking an x-ray of the dataset,
with the covariance coefficient equivalent to density, thus dark regions of
figure 3 correspond to volumes of high covariance. The position of the AC
planes in figure 4 may be identified through a comparison with figure 3. The
dark volumes between the AC planes are true correlations, generated only by
3BCE channels, identified in part by average TOF spectra displayed parallel
to the $\mathit{x}$,~$\mathit{y}$,~$\mathit{z}$ axes. The labels on the
average TOF spectra adhere to the peak assignments in table I.

An example of a 3BCE correlation between peaks 1 (C$^{3+}$), 5 (O$^{2+}_{F}$
/ S$^{4+}_{F}$) and 6 (O$^{2+}_{B}$ / S$^{4+}_{B}$) is identified in figure
4 by the intersection of three semi-transparent planes superimposed in the
3DCM. This correlation therefore represents the (2, 3, 4) channel, where the
ion order has been changed to agree with the ($\mathit{m}$, $\mathit{n}$, $%
\mathit{p}$) convention discussed earlier. However, the KER associated with
this channel cannot be measured, as there is uncertainty due to the
charge-to-mass degeneracy as to whether axes $\mathit{y}$ and $\mathit{z}$
represent O$^{2+}_{F}$ and S$^{4+}_{F}$ respectively or vice versa: the
solution of this problem will be presented later in this section.

\begin{figure}[tbp]
\includegraphics[width=235pt]{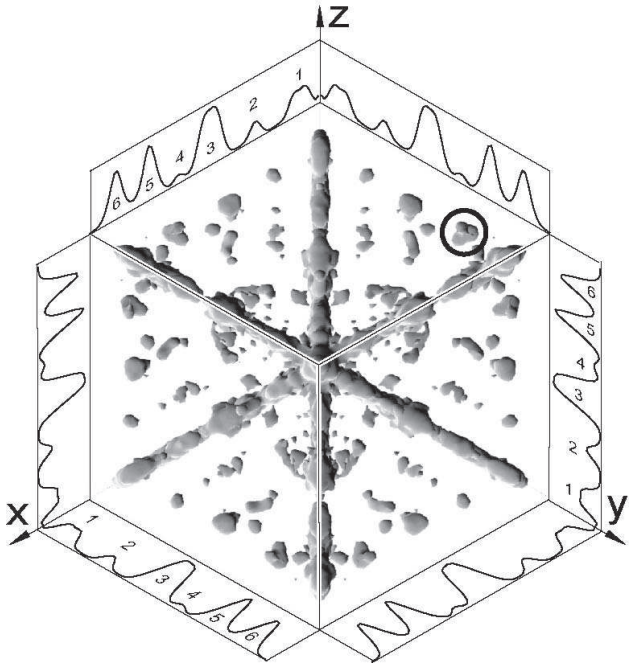}
\caption{Iso-surface visualization of the OCS three-dimensional covariance
map, where the complication of the dataset by the presence of the three
autocorrelation planes has been almost negated by viewing the map along the
direction $\mathit{x}$ = $\mathit{y}$ = $\mathit{z}$. Again, the peaks are
identified by comparison with table I. The circled region corresponds to the
three-body correlation forming the (2, 3, 4) channel.}
\label{fig5}
\end{figure}

Figure 5 is an iso-surface representation of the 3DCM, an
iso-surface being the 3D equivalent of a contour in 2D. The image
plane in figure 5 is defined normal to the line $\mathit{x}$ =
$\mathit{y}$ = $\mathit{z}$, dramatically reducing the impact of the
AC planes, as each of the AC planes $\mathit{x}$ = $\mathit{y}$,
$\mathit{x}$ = $\mathit{z}$ and $\mathit{y}$ = $\mathit{z}$ are
normal to the image plane. A series of iso-surfaces is used to
examine the 3DCM, thus identifying the 3BCE channels, where true
correlations are identified in the regions between the AC planes by
adjusting the iso-surface threshold. The C$_{3}$($\mathit{x}$,
$\mathit{y}$, $\mathit{z}$) volume is investigated by varying the
surface threshold and noting local maxima in the regions bounded by
the AC planes, corresponding to 3BCE channels. The primary benefit
of 3D visualization is that by moving the viewpoint from parallel to
the $\mathit{x}$, $\mathit{y}$ or $\mathit{z}$ axes, an analogue
to taking a slice, to a point on the line $\mathit{x}$ = $\mathit{y}$ = $%
\mathit{z}$ simplifies the system considerably. Furthermore, by examining
volumes of the map rather than slices, a more accurate identification may be
carried out, effectively removing the AC planes from the dataset and thus
making the volumes of interest immediately accessible.

\begin{table}[tbp]
\caption{Kinetic energy release (KER) associated with all 2BCE and
3BCE channels as identified from the 2DCM and 3DCM. The r$_{OC}$ and
r$_{CS}$ bond lengths are modeled using a Monte Carlo method, and
the channel strength is estimated by integrating the covariance
volumes.}
\label{tab:table2}%
\begin{ruledtabular}
\begin{tabular}{ccccc}
Channel &KER (eV) &{\it r}$_{OC}$ (\AA)&{\it r}$_{CS}$
(\AA)&Strength
(arb.)\\
\hline
(1,1) &4.1 &- &2.9 &0.5\\
(1,2) &7.6 &- &3.2 &0.2\\
(1,1,1)a &12.3 &2.5 &3.5 &0.7\\
(1,1,1)b &14.9 &2.0 &3.0 &1.0\\
(1,1,2) &17.9 &2.6 &3.7 &1.5\\
(1,2,2) &34 &2.3 &3.5 &2.0\\
(1,2,3) &38.9 &2.8 &3.9 &1.0\\
(2,2,2) &46.3 &2.6 &3.8 &1.5 \\
(1,2,4) &48.3 &2.5 &4.1 &1.0 \\
(2,2,3) &55.8 &2.7 &4.0 &3.0 \\
(2.3.2) &59.6 &2.8 &4.2 &2.5 \\
(1,3,4) &66 &2.6 &4.2 &1.0 \\
(2,2,4) &63 &2.9 &4.3 &4.0 \\
(2,3,3) &73  &3.0 &4.1 &4.0 \\
(2,3,4) &81 &3.1 &4.5 &5.0 \\
(3,2,4) &78 &3.2 &4.4 &2.0 \\
(3,3,4) &96 &3.3 &4.9 &4.0 \\
(3,3,5) &105 &3.4 &5.2 &2.0\\
\end{tabular}
\end{ruledtabular}
\end{table}

The KER associated with each of the correlations identified is calculated
from the time difference between the centre of the correlation volume and
the zero kinetic energy point for each ion charge-to-mass ratio parallel to
the $\mathit{x}$,~$\mathit{y}$,~$\mathit{z}$ axes. The total KER associated
with each correlation volume is presented in table II, along with an estimate
of the relative strength of the channel in the range 0 to 5. The detected
channel strength is gauged by integrating the correlation volumes. Returning
to identifying the correlations highlighted in figures 4 and 5, by
converting the KER (eV) to momentum, $\mathit{p}$ (amu ms$^{-1}$) for all
combinations of ion identities possible from table I, a straightforward
application of the conservation-of-momentum parallel to the detector axis
allows the correct bond length assignment to be made. An incorrect
assignment of the 3DCM correlation volume results in a net momentum
imbalance, whereas a correct assignment leads to zero net momentum. This
process is repeated for each correlation volume in figures 4 and 5, and for
all possible ion assignments for a particular group of three peaks, thus the
3BCE channels are unambiguously identified. The final assignments are
presented in table II.

In the case of a diatomic molecule Coulomb exploding, it is trivial to
estimate the bond length at the point of explosion \cite{ref19}. However, in
the case of 3BCE of OCS, which, as apparent from table II, is heavily
predisposed to charge-asymmetric Coulomb explosion, such a calculation is
not straightforward. To address this, the following technique was applied.
The bend angle ($\theta $) distribution between $\mathit{r}_{\text{OC}}$ and
$\mathit{r}_{\text{CS}}$ (see figure 1) for identical laser conditions is
known \cite{ref7}, and using the Monte-Carlo software developed as part of
our IMI technique \cite{ref4,ref6,ref7}, the 3BCE of OCS is simulated in
momentum space for a range of geometries. The bond lengths reported in table
II are those in strongest agreement with the 3DCM presented in figures 4 and
5, with $\mathit{r}_{\text{OC}}$, $\mathit{r}_{\text{CS}}$ and $\theta $
bound by conservation of momentum. As discussed earlier these bond lengths
assume a coulomb explosion from a stationary molecule. Although this is
clearly not the case, previous work \cite{ref7} suggests that even for the
highest channels the discrepancy is around 10\% between this approximation
and the correct bond length, and so rather than embark on the iterative
process set out previously \cite{ref7} we will begin our analysis with these
approximate values. A number of general trends are observed in table II:
there is an overall increase in the KER with increasing channel order, an
observation common to Coulomb explosion studies \cite%
{ref4,ref7,ref9,ref11,ref12,ref13,ref15}, caused by the higher Coulomb
repulsion between the ionic constituents. Furthermore, the $\mathit{r}_{%
\text{OC}}$ and $\mathit{r}_{\text{CS}}$ bonds are generally asymmetric
throughout, and tend to increase with channel ionization, $\mathit{q}$ = ($%
\mathit{m}$ + $\mathit{n}$ + $\mathit{p}$). Importantly, two (1, 1, 1)
channels are observed in the 3DCM, labeled ($\mathit{a}$) and ($\mathit{b}$%
) in table II in order of increasing KER. There should clearly be a dynamic
coupling between the 3BCE channels presented in table II, however the
tabulation of the results is somewhat ambiguous.

\section{Discussion}

\begin{figure}[tbp]
\includegraphics[width=235pt]{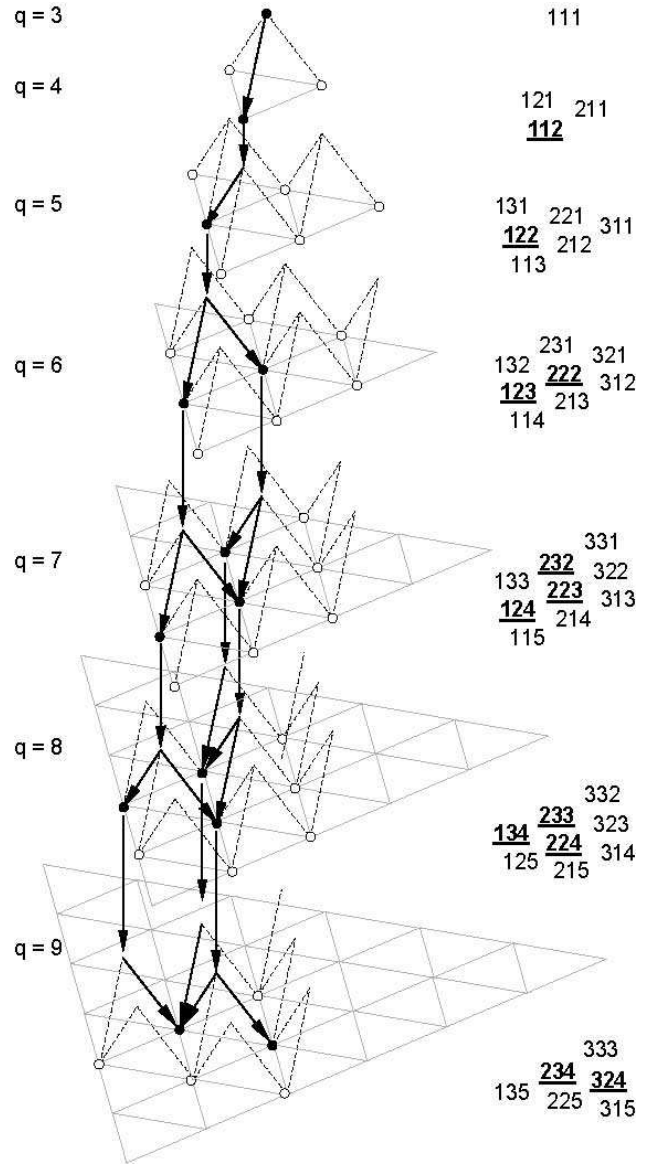}
\caption{General hierarchical ionization (GHI) diagram for the 3BCE
channels observed in the 3DCM as presented in table II. Open circles
represent numerically possible channels, solid circles represent
true 3BCE channels.} \label{fig6}
\end{figure}

To comprehend the possible pathways which the OCS molecule can take during
ionization we first make the assumption that the ionization channels which
we observe are the steps on an ionization ladder which finishes at (335). To
visualize this we have adapted the method of \cite{refH} into a 3D
hierarchical ionization topology. The general hierarchical ionization (GHI)
diagram for laser-OCS interaction is presented in figure 6, where the
ionization channels presented in table II, identified from the 3DCM presented
in figures 4 and 5 are indicated as nodes on a pyramidal framework. The
least-charged (lowest order) channel (1, 1, 1) is the node at the apex of
the pyramid, corresponding to a total charge, $\mathit{q}$ = 3. Progression
down the pyramid from the apex between the subsequent triangular grids
indicates ionization through the loss of one or more electrons. Each
intersection on a particular triangular grid uniquely represents a channel.
Three classes of node are introduced in figure 6: (1) each intersection
represents a numerically possible channel, (2) the open circles represent an
experimentally possible channel, and (3) the filled circles represent
observed 3BCE channels. The refinement to these three classes of node makes
the distinction between whether an ion may be generated, is observed in the
average time-of-flight spectrum or is observed as a true 3BCE channel,
following the application of conservation of momentum as discussed earlier.
The key to the right of figure 6 indicates experimentally possible channels
(regular type) and observed 3BCE channels (bold underlined). For example, on
the $\mathit{q}$ = 9 level, 28 channels are numerically possible,
illustrated by the large basic grid. However, the necessary component ions
are only observed for six channels (indicated in the key), and of these six,
only two triply-correlated channels are observed in the 3DCM for this level.
The vertical arrows in figure 6 indicate the transfer between \textit{q}%
-levels, the heads of which reach the vertices of three-sided secondary
pyramids on the next \textit{q}-level. At this point the overall charge of
the molecule is known, but the specific ionization channel is not. The
ionization routes emanate from the vertex of the secondary pyramid. In
choosing one of the three sides of this pyramid, the ionization channel is
specified. At this point, molecular ionization is treated as a single-site
sequential process. GHI is well illustrated by the transition between the
(2, 2, 3) channel on the $\mathit{q}$ = 7 level and the (2, 2, 4) and (2, 3,
3) channels on the $\mathit{q}$ = 8 level.

\begin{figure}[tbp]
\includegraphics[width=235pt]{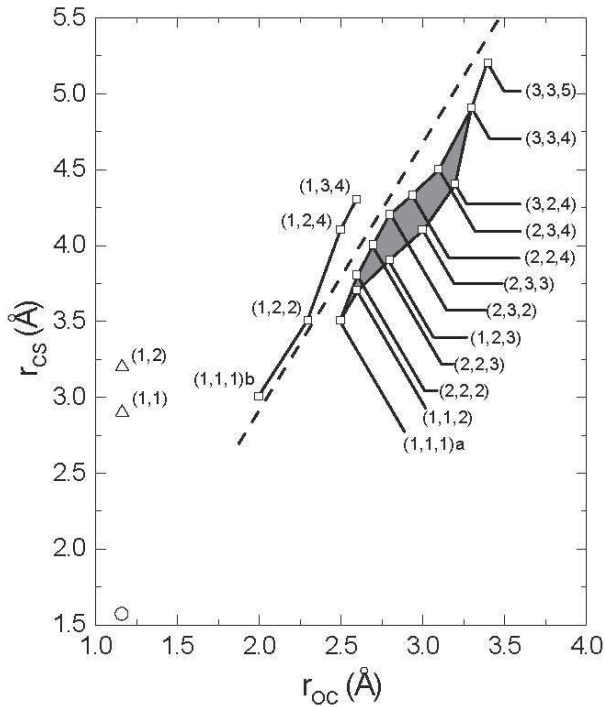}
\caption{The 3BCE channels identified from the 3DCM presented in ($\mathit{r}%
_{\text{OC}}$, $\mathit{r}_{\text{CS}}$) space, forming the second
stage in hierarchical ionization topology. The two groups of 3BCE
channels observed are separated by a dashed line for clarity.}
\label{fig7}
\end{figure}

The most striking feature of figure 6 is the apparent profusion of
ionization routes through the GHI diagram. We now turn to the second stage
of hierarchical ionization topology, and consider the influence of the
geometry of the exploding molecule as it develops through the GHI diagram.
Figure 7 shows the result of plotting in ($\mathit{r}_{\text{OC}}$, $\mathit{%
r}_{\text{CS}}$) space all of the ionization channels uniquely identified
using 3DCM. The first observation which is clear from figure 7 is that the
approximate bond lengths are not particularly constant as would be expected
from an enhanced ionization explanation, although on further inspection they
do form a group centered around extended bond lengths of $\mathit{r}_{\text{%
CS}}$=3.7au and $\mathit{r}_{\text{OC}}$=2.7au which would
correspond 2.5 and 2.25 times the equilibrium bond lengths which is
similar to the factor of 2, observed for CO$_{2}$ \cite{ref4}. The
second observation is that both bond lengths increase with charge
state but in an oscillating fashion (see shaded area figure)
\cite{ref7}. This increasing trend is similar to that observed
previously for CO$_{2}$ \cite{ref9}. In addition the oscillation is
accompanied by a set of channels somewhat offset from the others.
Specifically, the group of 12 3BCE channels, starting with (1, 1,
1)a and finishing at (3, 3, 5) contains the majority of channels,
which are generated at larger $\mathit{r}_{\text{OC}}$ and
$\mathit{r}_{\text{CS}}$
locations. The channels (1, 1, 1)b to (1, 3, 4), are generated at smaller $%
\mathit{r}_{\text{OC}}$ and $\mathit{r}_{\text{CS}}$ as compared to the main
group. The dashed diagonal line running from bottom left to top right serves
as a visual guide, separating the two regions of ($\mathit{r}_{\text{OC}}$, $%
\mathit{r}_{\text{CS}}$). For illustrative purposes three solid
lines link the ionization channels. At this point we must be very
careful about the conclusions we draw. It is tempting to see the
bond information, in table II and in figure 7, as being indicative
of those bond lengths occupied during the steps in the ionization
ladder illustrated in figure 6, and this is exactly the procedure
carried out by \cite{refG, refH, ref21}, although the bond lengths
are calculation differently to here in \cite{refG, refH}. But it is
important to realize that however accurate the bond length
measurement are, for a particular ionization channel say (223), it
does not mean that this ionization state uniquely occurs at the
measured bond lengths. A higher ionization state, say (233), formed
in a region of the focus of higher peak laser intensity, might
display the same bond lengths because of enhanced ionization. In
this case the molecule could still have passed through the (223)
state, but at shorter bond lengths. This is best illustrated for OCS
by figure 8 in \cite{ref7} which shows that although each channel
may ionize last at the critical distance the precursors to any given
channel will ionize at shorter bond lengths. These shorter bond
lengths are not measurable with a long laser pulse. This phenomenon
is consistent with the recent work using short (sub 7fs) laser
pulses \cite{refA, refB} which clearly shows ionization at bond
lengths less than the critical distance, if ionization is switched
off rapidly before the critical distance is reached. For pulses of
50fs or longer, it is even possible that bond lengths measured for
say the (233) channel could be shorter than those measured for the
(223) channel but that the (223) state is a precursor of the (233)
channel. Two 2BCE channels are also included in figure 7. It is
assumed that the OC bond has not expanded during
2BCE, thus 2BCE can be thought of as starting from the equilibrium geometry (%
$\circ $, figure 7), then the CS bond expands while the OC bond is
constant, undergoing ionization to the (1, 1) or (1,2) channels
($\triangle $, figure 7). \ Furthermore the dissociation of OCS into
both (1,1) and (1,2) will be quite nonCoulombic and so the
approximate bond lengths given in figure 7 are expected to be an
overestimate. It is even likely that these channels dissociate from
equilibrium.

Figure 7 makes a distinction between two regions in ($\mathit{r}_{\text{OC}}$%
, $\mathit{r}_{\text{CS}}$) space. In the case of the four channels to the
left of the dashed diagonal line, multiple ionization only occurs from the
carbon or sulphur sites. The charge on the oxygen site remains 1, even
though the carbon is observed to loose up to 2 more electrons and the
sulphur up to 3 more. In contrast, the channels to the right of the dashed
line in figure 7 are observed to Coulomb explode with levels of ionization
loosely bound by the ionization potentials of the constituent atoms. At this
point we tentatively proposed that the (111)b channel and the asymmetric
channels are associated and form a different ionization ladder to the
channels to the right of the line.

\begin{figure}[tbp]
\includegraphics[width=235pt]{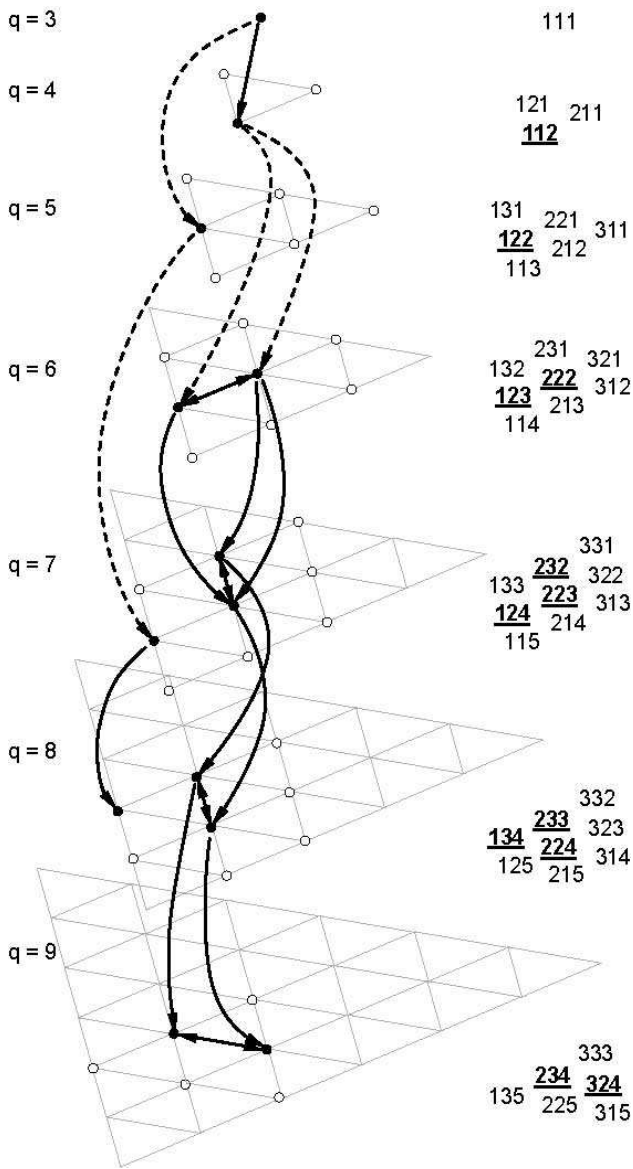}
\caption{Refined hierarchical ionization (RHI) diagram for the 3BCE
channels observed in the 3DCM after simplification as described in
the text. The solid arrows indicate single ionization, the dashed
arrows indicate double ionization.} \label{fig8}
\end{figure}

Figure 8 illustrates the refined hierarchical ionization (RHI) diagram for
the laser-OCS interaction. Visually, figure 8 appears simpler and better
defined in comparison to figure 6 as there are fewer connections between
nodes; however, the processes implied are significant. The three 3BCE
asymmetric channels which are observed after the (111)$b$ channel are
generated through two stages of double ionization. Dashed lines in figure 8
indicate a double ionization step. A pair of double ionization steps also
occur after the (111)$a$ channel on route to (1,2,3) and (2,2,2) from
(1,1,2).

\begin{figure}[tbp]
\includegraphics[width=235pt]{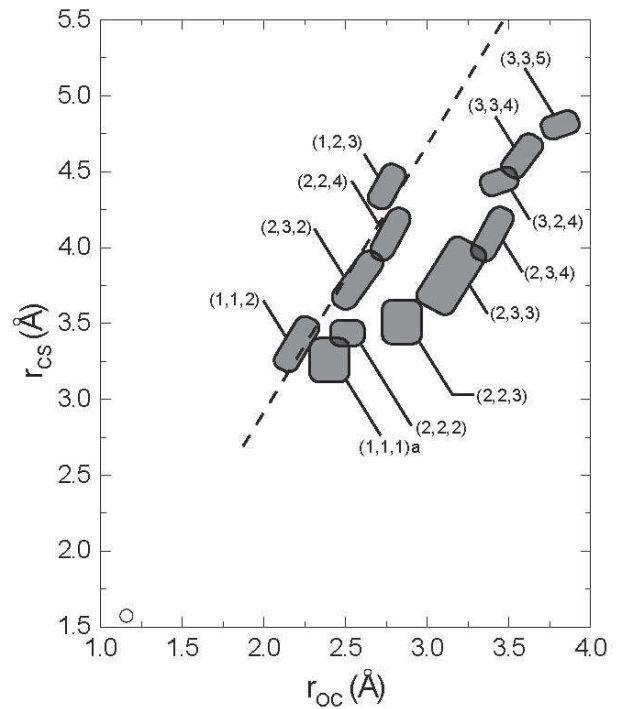}
\caption{Simulation of ionization to the right of dashed the line in figure
7, using the modified enhanced ionization (EI) model, described in the text.
An analogous simulation of channels to the left of the line was
unsuccessful. }
\label{fig9}
\end{figure}

We now attempt to reproduce the results to the right of the dashed line in
figure 7 using the results of a series of EI calculations. Unlike the
previous calculations presented in \cite{ref7},
the current calculations allow both bonds to expand independently, thus for
each ($\mathit{m}$, $\mathit{n}$, $\mathit{p}$) channel an appearance
intensity surface is created in ($\mathit{r}_{\text{OC}}$, $\mathit{r}_{%
\text{CS}}$) space. This large calculation is possible by only calculating
the effective Coulomb potential (Coulombic and laser field) along $\mathit{r%
}_{\text{OC}}$ and $\mathit{r}_{\text{CS}}$, as it has been found that this
potential is minimized along these axes, hence EI will depend only on these
conditions. This simplification only applies for near linear molecules (in
this case $\theta $ = 170 degrees) parallel to the laser polarization
direction, reasonable in this case given the observations of Sanderson $%
\textit{et al}$ \cite{ref7}. To best replicate the laser conditions
responsible for the observations of figure 7, a wide range of laser
intensities are modeled, as the laser intensity defines which
ionization channel occurs. Each channel occupies a region of
($\mathit{r}_{\text{OC}}$, $\mathit{r}_{\text{CS}}$) space,
indicated by the shaded areas in figure 9. As is apparent from
comparison between figures 7 and 9, there is reasonable
agreement between the measured and predicted distribution in ($\mathit{r}_{%
\text{OC}}$, $\mathit{r}_{\text{CS}}$), especially given the number of
ionization channels involved.

As a development of the original treatment of the Coulomb explosion of
diatomic molecules \cite{ref19}, the reasonable agreement between the figures 7
and 9 is only possible following two major modifications to the model.
Firstly, using the accepted ionization potentials of the constituent atoms,
the appearance intensity minimum in ($\mathit{r}_{\text{OC}}$, $\mathit{r}_{%
\text{CS}}$) space (the critical geometry) is considerably larger than
equilibrium for low \textit{q} channels, and moves to lower $\mathit{r}_{%
\text{OC}}$, $\mathit{r}_{\text{CS}}$ with increasing total charge $\mathit{q%
}$. So, even before taking into account the laser pulse trajectory \cite%
{ref19}, the trend observed in figure 7 cannot be replicated within the
unmodified EI model. By decreasing the atomic ionization potentials with
increasing charge, the appearance intensity minimum is varied to give a far
better agreement with the present observations. The best-fit ionization
potential $\mathit{I}_{p}^{~\prime }$ = $\mathit{k}$ $\mathit{I}_{p}$ where $%
\mathit{I}_{p}$ is the accepted ionization potential and $\mathit{k}$ is the
scaling parameter given by $\mathit{k}$ = - 0.1($\mathit{m}$, $\mathit{n}$)
+ 1 or $\mathit{k}$ = - 0.05$\mathit{p}$ + 0.95 depending on the channel ($%
\mathit{m}$, $\mathit{n}$, $\mathit{p}$). Secondly, calculating the laser
trajectory as the molecule expands reveals a dramatic underestimate of the
bond extension for all channels, for example, the unmodified model predicts
the (3,3,5) channel Coulomb explodes from $\mathit{r}_{\text{OC}}$ = 2.5 \AA
, $\mathit{r}_{\text{CS}}$ = 3.0 \AA . Careful investigation of the
intersections between the appearance surfaces and the laser trajectories
over a wide intensity reveals the low $\mathit{q}$ channels ionizing with
near equilibrium bonds, thus limiting the expansion possible in the higher $%
\mathit{q}$ channels. The behaviour of the low $\mathit{q}$ channels
indicates that molecular expansion is not purely Coulombic, a simple
solution to the overestimate of the interaction is to reduce the charge of
the ions by assigning fractional charges, adjusted to give a best fit to
figure 7. It was found that the low $\mathit{q}$ channels, i.e. (111),
required charge suppression $\approx $ 0.7, whereas the high $\mathit{q}$
channels, i.e. (335) needed no charge suppression, intermediate channels
exhibit a linear relationship between these limits.

The modifications made to the enhanced ionization model do have
valid physical basis. Firstly the modifications to $q$ effectively
introduce a bonding aspect to the low charge states and make them
less Coulombic. This deviation from pure Coulomb interaction for low
charge states is well known in such molecules as Cl$_{2}$
\cite{refK} and decreases with charge state. The small modification
to the atomic IP's can be related to the ionization potential of the
molecule, or the gap between the molecular interaction potentials.
For low charge states the gap is larger than it would be for purely
Coulombic interactions because the bonding character of the
potential, represented by a dip below the Coulombic curve, reduces
in magnitude as charge state increases. With these modifications it
is possible to represent the general features of the majority of the
ionization channels, in a way which is self consistent throughout
and this is a significant indication of the overall validity of the
enhanced ionization model. However it is equally significant that we
can not represent all final ionization channels with this approach.
The (111)$b$ channel and the other channels to the left of the
dashed line in figure 7 can not be accommodated by this treatment.

In an effort to account for the remaining channels we should begin by
explaining the two distinct energy groups for the (111) channel. Firstly it
is sensible to consider enhanced ionization as it has been reasonably
successful thus far. In the enhanced ionization picture bond expansion
towards the critical distance is require before CE is complete. This would
seem to imply the existence of two critical distances for the (111) channel.
Although a secondary minimum was found in the ionization threshold curve
calculated for the (334) channel previously \cite{ref7} none was found for the
(111) channel. Even if a double structure does exist for an ionization
threshold curve it does not easily give rise to double valued bond lengths,
because if an ionization trajectory crosses the threshold curve once, the
molecule will ionize and there is no advantage to crossing more than once. A
second possibility is the presence of two distinct energetic channels in the
break up of OCS$^{3+}$, this might be more likely as we observe stable OCS$%
^{3+}$ in the time of light spectrum of this molecule, a quite
unusual observation for triatomic molecules; it is not seen for
CO$_{2}$ or N$_{2}$O for instance. It is possible that a long lived
metastable trication may completely dissociate after the laser pulse
from equilibrium geometry. The (111)$b$ channel appears to have
stretched from equilibrium but this could be deceptive just as
$\mathit{r}_{\text{CS}}$ appears to have stretched in the (11) and
(12) channel this may just be the effect of a non Coulombic
dissociation channel. One problem with this possibility is that
(111)$b$ appears to be precursor of the other channels such a s
(122) and this could not take place after the laser pulse. A third
possibility is that two distinctive dissociation channels in the
OCS$^{2+}$ molecule exist, which give rise to different degrees of
atomic excitation. The more highly excited molecule would ionize to
the (111) channel at a lower laser intensity, this phenomenon has
been observed for metastable argon ions \cite{refL}. A consequence
of this would be to shift the critical distance to lower bond
length, in fact two curves would exist for the two different
molecular states with the same charge. It should be notes that two
distinct kinetic energy groups have been observed for the S fragment
in the CO+S dissociation of OCS after 233nm absorption \cite{ref33}.
This phenomenon could not be predicted from the simple classical
mode, as it would require two distinct sets of atomic ionization
potentials to exist. Excited molecules formed in high intensity
regions would reach the (111) channel earlier in the laser pulse and
this may be why asymmetric double ionization is possible after the
(111)$b$ channel is formed, as multi electronic asymmetric processes
occur at bond lengths closer to equilibrium than the critical
distance \cite{refH}. If asymmetric ionization is preferred at
shorter bond lengths then only the state which gives rise to the
(111)$b$ channel can make the triply ionized molecule at
sufficiently small bond length to allow further asymmetric
multi-electronic processes to occur. The state which give rise to
the (111)$a$ channel does not do so until after the molecule has
passed through this bond length range.

\section{Conclusion}

The kinetic energy release during two- and three-body Coulomb explosion of
carbonyl sulfide has been measured using three-dimensional covariance
mapping. For the first time in a triatomic molecule two distinct energy
groups have been identified for the same dissociation channel. A 3D
hierarchical ionization topology has been introduced to help visualize the
multitude of possible ionization pathways and a Classical trajectory
simulation using a simple over the barrier enhanced ionization model has
been used in an attempt to reproduce the observed bond lengths. The
distribution of most bond lengths are reasonably predicted when small
modification to the enhanced ionization model are introduced. The presence
of the two (111) channels and highly asymmetric channels however can not be
modeled using the simple approach.

In conclusion, OCS has proven to be a useful subject of study in the quest
to understand the ionization mechanisms of small molecules in an intense
laser pulse. The results indicate the need to develop an enhanced ionization
model which incorporates molecular excitation and the possibility of
multi-electronic processes in order to be able to fully predict the
ionization process. In addition the need for further experimental study is
indicated, in order to better understand the interplay between the
ionization routes indicated by the current work, utilization of sub 10 femtosecond
pulses would enable probing within the critical distance and full
coincidence techniques would establish the link between geometry and
ionization pathway.

\begin{acknowledgments}
This work is supported by the Engineering and Physical Sciences Research
Council, and NSERC.
\end{acknowledgments}

\bibliographystyle{apsrev}
\bibliography{ocspap2}

\end{document}